\documentclass[12pt]{article}
\def\be{\begin{equation}}\def\ee{\end{equation}}

\makeatletter\ifcase\@ptsize\font\teneufm=eufm10
\font\seveneufm=eufm7\font\fiveeufm=eufm5
\font\teneusm=eusm10\font\seveneusm=eusm7
\font\fiveeusm=eusm5\or\font\teneufm=eufm10 scaled
\magstephalf\font\seveneufm=eufm7\font\fiveeufm=eufm5
\font\teneusm=eusm10 scaled\magstephalf
\font\seveneusm=eusm7\font\fiveeusm=eusm5\or
\font\teneufm=eufm10 scaled\magstep1\font\seveneufm=eufm7
\font\fiveeufm=eufm5\font\teneusm=eusm10 scaled\magstep1
\font\seveneusm=eusm7\font\fiveeusm=eusm5\fi

\newfam\eufmfam\newfam\eusmfam\textfont\eufmfam=\teneufm
\scriptfont\eufmfam=\seveneufm
\scriptscriptfont\eufmfam=\fiveeufm
\textfont\eusmfam=\teneusm\scriptfont\eusmfam=\seveneusm
\scriptscriptfont\eusmfam=\fiveeusm

\def\frak{\ifmmode\let\next\frak@\else
\def\next{\errmessage{Use\string\frak\space only in math
mode}}\fi\next}\def\frak@#1{{\frak@@{#1}}}
\def\frak@@#1{\fam\eufmfam#1}
\def\sh{\ifmmode\let\next\sh@\else
\def\next{\errmessage{Use\string\sh\space only in math
mode}}\fi\next}\def\sh@#1{{\sh@@{#1}}}
\def\sh@@#1{\fam\eusmfam#1}

\ifcase\@ptsize\font\tenmsa=msam10\font\sevenmsa=msam7
\font\fivemsa=msam5\font\tenmsb=msbm10
\font\sevenmsb=msbm7\font\fivemsb=msbm5\or
\font\tenmsa=msam10 scaled\magstephalf
\font\sevenmsa=msam7\font\fivemsa=msam5
\font\tenmsb=msbm10 scaled\magstephalf
\font\sevenmsb=msbm7\font\fivemsb=msbm5\or
\font\tenmsa=msam10 scaled\magstep1\font\sevenmsa=msam7
\font\fivemsa=msam5\font\tenmsb=msbm10 scaled\magstep1
\font\sevenmsb=msbm7\font\fivemsb=msbm5\fi

\newfam\msafam\newfam\msbfam\textfont\msafam=\tenmsa
\scriptfont\msafam=\sevenmsa
\scriptscriptfont\msafam=\fivemsa\textfont\msbfam=\tenmsb
\scriptfont\msbfam=\sevenmsb
\scriptscriptfont\msbfam=\fivemsb

\def\Bbb{\ifmmode\let\next\Bbb@\else
\def\next{\errmessage{Use\string\Bbb\space only in math
mode}}\fi\next}\def\Bbb@#1{{\Bbb@@{#1}}}
\def\Bbb@@#1{\fam\msbfam#1}\def\hexnumber@#1{\ifnum#1<10
\number#1\else\ifnum#1=10 A\else\ifnum#1=11
B\else\ifnum#1=12 C\else\ifnum#1=13 D\else\ifnum#1=14
E\else\ifnum#1=15 F\fi\fi\fi\fi\fi\fi\fi}
\def\msa@{\hexnumber@\msafam}\def\msb@{\hexnumber@\msbfam}
\mathchardef\square="0\msa@03

\makeatother
\newcommand\RR{{\Bbb R}}
\newcommand{\CC}{{\Bbb C}}

\begin{document}

\def\title#1{\centerline{\huge{#1}}}
\def\author#1{\centerline{\large{#1}}}
\def\address#1{\centerline{\it #1}}
\def\ack{{\bf Acknowledgments}$\quad$}
\def\Bibliography{\begin{thebibliography}}
\def\endbib{\end{thebibliography}}


\begin{titlepage}


\title{The Cocycle of the Quantum HJ Equation}

\vspace{.333cm}

\title{and the Stress Tensor of CFT}
\vspace{1.5cm}
\author{Marco Matone}
\vspace{.3cm}
\address{
Department of Physics ``G. Galilei'' -- Istituto Nazionale di
Fisica Nucleare}
\address{University of Padova, Via Marzolo, 8 -- 35131 Padova, Italy}

\begin{abstract} \noindent
We consider two theorems formulated in the derivation of the
Quantum Hamilton--Jacobi Equation from the EP. The first one
concerns the proof that the cocycle condition uniquely defines the
Schwarzian derivative. This is equivalent to show that the
infinitesimal variation of the stress tensor ``exponentiates" to
the Schwarzian derivative. The cocycle condition naturally defines
the higher dimensional version of the Schwarzian derivative
suggesting a role in the transformation properties of the stress
tensor in higher dimensional CFT. The other theorem shows that
energy quantization is a direct consequence of the existence of
the quantum Hamilton--Jacobi equation under duality
transformations as implied by the EP.
\end{abstract}

\end{titlepage}

\section{HJ Equation and Coordinate Transformations}

Let us consider the Hamilton--Jacobi (HJ) equation for a one
dimensional system. This is obtained by considering the canonical
transformation $(q,p)\to (Q,P)$ leading to a vanishing
Hamiltonian$$ \tilde H=0\ . $$ The old and new momenta are
expressed in terms of the generating function of such a
transformation, the Hamilton's principal function
$$p={\partial{\cal S}^{cl}\over\partial q}\ ,\qquad
P=cnst=-{\partial{\cal S}^{cl}\over
\partial Q}{|_{Q=cnst}}\ ,
$$ that satisfies the classical HJ equation $$
H\left(q,p={\partial{\cal S}^{cl}\over\partial q},t\right)+
{\partial{\cal S}^{cl}\over\partial t}=0\ . $$ In the case of a time
independent potential the time dependence in the Hamilton's
principal function ${\cal S}^{cl}$ is linear, that is
$$ {\cal S}^{cl}(q,Q,t)={\cal S}_0^{cl}(q,Q)-Et\ , $$ where $E$ denotes the energy of the
stationary state. It follows that ${\cal S}_0^{cl}$, called Hamilton's
characteristic function or reduced action, satisfies the Classical
Stationary HJ Equation (CSHJE) $$
H\left(q,p={\partial{\cal S}_0^{cl}\over\partial q}\right)-E=0\ , $$
that is (${\cal W}(q)\equiv V(q)-E$) $$
{1\over2m}\left({\partial{\cal S}_0^{cl}\over\partial q}\right)^2+{\cal W}=0\ .
$$

Following \cite{1} we now consider a similar question to that
leading to the CSHJE but starting with $$
p={\partial{\cal S}_0^{cl}\over\partial q}\ ,
$$ rather than with $p$ and $q$ considered as independent
variables. More precisely, given a one dimensional system, with
time--independent potential (the higher dimensional time--dependent
case is considered in \cite{BFM}) we look for the coordinate
transformation $q\to q_0$ such that \be {\cal
S}_0^{cl}(q)\qquad\stackrel{Coord.\;Transf.}{\longleftrightarrow}\qquad\tilde
{\cal S}_0^{cl\,0}(q_0)\ , \label{staccorello2}\ee with $\tilde{\cal
S}_0^{ cl\,0}(q_0)$ denoting the reduced action of the system with
vanishing Hamiltonian. Note that in (\ref{staccorello2}) we required
that this transformation be an invertible one. This is an important
point since by compositions of the maps it follows that if for each
system there is a coordinate transformation leading to the trivial
state, then even two arbitrary systems are equivalent under
coordinate transformations. Imposing this apparently harmless
analogy immediately leads to rather peculiar properties of Classical
Mechanics (CM). First, it is clear that such an equivalence
principle cannot be satisfied in CM, in other words given two
arbitrary systems $a$ and $b$, the condition \be{\cal
S}_0^{cl\,b}(q_b)={\cal S}_0^{cl\,a}(q_a)\ , \label{staccc}\ee
cannot be generally satisfied. In particular, since
$$\tilde{\cal S}_0^{cl\,0}(q_0)=cnst\ ,$$ it is clear that
(\ref{staccorello2}) is a degenerate transformation. However, in
principle, by itself the failure of (\ref{staccc}) for arbitrary
systems would be a possible natural property. Nevertheless, a more
careful analysis shows that such a failure is strictly dependent
on the choice of the reference frame. This is immediately seen by
considering two free particles of mass $m_a$ and $m_b$ moving with
relative velocity $v$. For an observer at rest with respect to the
particle $a$ the two reduced actions are
$$ {\cal S}_0^{cl\,a}(q_a)=cnst\ ,\qquad{\cal S}_0^{cl\,b}(q_b)=m_bvq_b\ . $$ It
is clear that there is no way to have an equivalence under
coordinate transformations by setting ${\cal S}_0^{cl\,b}(q_b)={\cal
S}_0^{cl\,a}(q_a)$. This means that at the level of the reduced
action there is no coordinate transformation making the two systems
equivalent. However, note that this coordinate transformation exists
if we consider the same problem described by an observer in a frame
in which both particles have a non--vanishing velocity so that the
two particles are described by non--constant reduced actions.
Therefore, in CM, it is possible to connect different systems by a
coordinate transformation except in the case in which one of the
systems is described by a constant reduced action. This means that
in CM equivalence under coordinate transformations is frame
dependent. In particular, in the CSHJE description there is a
distinguished frame. This seems peculiar as on general grounds what
is equivalent under coordinate transformations in all frames should
remain so even in the one at rest.

\section{The Equivalence Postulate}

The above investigation already suggests that the concept of point
particle itself cannot be consistent with the equivalence under
coordinate transformations. In particular, it suggests that the
system where a particle is at rest does not exist at all. If this
would be the case, then the above critical situation would not
occur simply because the reduced action is never a constant. This
should reflect in two main features. First the classical concept
of point particle should be reconsidered, secondly the CSHJE
should be modified accordingly. A natural suggestion would be to
consider particles as a kind of string with a lower bound on the
vibrating modes in such a way that there is no way to define a
system where the particle is at rest. It should be observed that
this kind of string may differ from the standard one, rather its
nature may be related to the fact that in general relativity is
impossible to define the concept of relative stability of a system
of particles.

We now start imposing the equivalence under coordinate
transformations. The key point is to consider, like in general
relativity, the (analogous of the) reduced action as a scalar
field under coordinate transformations.

We postulate that for any pair of one--particle states there
exists a field $\mathcal{S}_0$ such that \be
\mathcal{S}_0^b(q_b)=\mathcal{S}_0^a(q_a)\ , \label{abinitio}\ee is
well defined. We also require that in a suitable limit
$\mathcal{S}_0$ reduces to $\mathcal{S}_0^{cl}$.
Eq.(\ref{abinitio}) can be considered as the scalar hypothesis.
Since the conjugate momentum is defined by $$
p_i={\partial\over\partial q^i} \mathcal{S}_0(q)\ ,$$ it follows
by (\ref{abinitio}) that the conjugate momenta $p^a$ and $p^b$ are
related by a coordinate transformation \be
p^b_i=\Lambda_i^{\,j}p^a_j\ , \label{dddff}\ee where
$\Lambda_i^{\,j}={\partial q_a^j/\partial q_b^i}$. Note that we
have the invariant \be p_i^bdq_b^i=p_i^adq_a^i\ . \label{invva}\ee
Since (\ref{abinitio}) holds for any pair of one--particle states,
we have ${\rm Det}\, \Lambda(q)\neq 0$, $\forall q$.

The scalar hypothesis (\ref{abinitio}) implies that two
one--particle states are always connected by a coordinate
transformation, for such a reason we may equivalently consider
(\ref{abinitio}) as imposing an Equivalence Postulate (EP). In
particular, while in arbitrary dimension the coordinate
transformation is given by imposing (\ref{dddff}), in the one
dimensional case the scalar hypothesis immediately leads to
$$q_b={{\cal S}_0^b}^{-1}\circ{\cal S}_0^a(q_a) \ . $$

We now consider the consequences of the EP (\ref{abinitio}). Let
us denote by ${\cal H}$ the space of all possible ${\cal W}\equiv V-E$.
We also call $v$--transformations the ones leading from a system
to another. Eq.(\ref{abinitio}) is equivalent to require that

\vspace{.333cm}

\noindent {\it For each pair ${\cal W}^a,{\cal W}^b\in{\cal H}$, there is a
$v$--transformation such that} \be
{\cal W}^a(q)\longrightarrow{{\cal W}^a}^v(q^v)={\cal W}^b(q^v)\ .
\label{equivalence}\ee This implies that there always exists the
trivializing coordinate $q_0$ for which
${\cal W}(q)\longrightarrow{\cal W}^0(q_0)$, where $$ {\cal W}^0(q_0)\equiv 0\ .
$$ In particular, since the inverse transformation should exist as
well, it is clear that the trivializing transformation should be
locally invertible. We will also see that since classically ${\cal W}^0$
is a fixed point, implementation of (\ref{equivalence}) requires
that ${\cal W}$ states transform inhomogeneously.

The fact that the EP cannot be consistently implemented in CM is
true in any dimension. To show this let us consider the coordinate
transformation induced by the identification \be {\cal S}_0^{cl\,
v}(q^v)={\cal S}_0^{cl}(q)\, . \label{pcowcOII}\ee Then note that the CSHJE
\be
{1\over2m}\sum_{k=1}^D({\partial_{q_k}{\cal S}_0^{cl}(q)})^2+{\cal W}(q)=0\ ,
\label{012}\ee provides a correspondence between ${\cal W}$ and ${\cal S}_0^{cl}$
that we can use to fix, by consistency, the transformation
properties of ${\cal W}$ induced by that of ${\cal S}_0^{cl}$. In particular,
since ${\cal S}_0^{cl\, v}(q^v)$ must satisfy the CSHJE \be
{1\over2m}\sum_{k=1}^D(\partial_{q^{k\,
v}}{\cal S}_0^{cl\,v}(q^v))^2+{\cal W}^v(q^v)=0\ , \label{zummoloa30}\ee by
(\ref{pcowcOII}) we have \be
{\partial{\cal S}_0^{cl\,v}(q^v)\over\partial q^{k\, v}}=\Lambda_k^{\,
i}{\partial{\cal S}_0^{cl}(q)\over\partial q^i}\ . \label{insomma}\ee Let us
set $(p^v|p)={p^t\Lambda^t\Lambda p/ p^tp}$. By
(\ref{012})--(\ref{insomma}), we have
${\cal W}(q)\longrightarrow{\cal W}^v(q^v)=(p^v|p){\cal W}(q)$, so that $$
{\cal W}^0(q_0)\longrightarrow{\cal W}^v(q^v)=(p^v|p^0){\cal W}^0(q_0)=0\ . $$ Thus
we have \cite{1}

\vspace{.333cm}

\noindent {\it ${\cal W}$ states transform as quadratic differentials
under classical $v$--maps. It follows that ${\cal W}^0$ is a fixed point
in ${\cal H}$. Equivalently, in CM the space ${\cal H}$ cannot be reduced to a
point upon factorization by the classical $v$--transformations.
Hence, the EP (\ref{equivalence}) cannot be consistently
implemented in CM. This can be seen as the impossibility of
implementing covariance of CM under the coordinate transformation
defined by (\ref{pcowcOII}).}

\vspace{.333cm}

It is therefore clear that in order to implement the EP we have to
deform the CSHJE. As we will see, this requirement will determine
the equation for ${\cal S}_0$.

In Ref.\cite{1} the function ${\cal T}_0(p)$, defined as the Legendre
transform of the reduced action, was introduced $$
{\cal T}_0(p)=q^kp_k-{\cal S}_0(q),\qquad{\cal S}_0(q)=p_kq^k-{\cal T}_0(p)\ . $$ While
${\cal S}_0(q)$ is the momentum generating function, its Legendre dual
${\cal T}_0(p)$ is the coordinate generating function $$
p_k={\partial{\cal S}_0\over\partial q_k},\qquad
q_k={\partial{\cal T}_0\over\partial p_k}\ . $$ Note that adding a
constant to ${\cal S}_0$ does not change the dynamics. Then, the most
general differential equation ${\cal S}_0$ should satisfy has the
structure \be {\cal F}(\nabla{\cal S}_0,\Delta{\cal S}_0,\ldots)=0\ .
\label{traslazione}\ee \noindent Let us write down
Eq.(\ref{traslazione}) in the general form $$
{1\over2m}\sum_{k=1}^D({\partial_{q^k}{\cal S}_0(q)})^2+{\cal W}(q)+Q(q)=0\ .
$$ The transformation properties of ${\cal W}+Q$ under the $v$--maps are
determined by the transformed equation \be
{1\over2m}\sum_{k=1}^D(\partial_{q^{k\,
v}}{\cal S}_0^v(q^v))^2/2m+{\cal W}^v(q^v)+Q^v(q^v)=0\ , \label{labella998}\ee so
that \be {\cal W}^v(q^v)+Q^v(q^v)=(p^v|p)\left[{\cal W}(q)+Q(q)\right]\ .
\label{yyyxxaa10bbbb}\ee

A basic guidance in deriving the differential equation for ${\cal S}_0$
is that in some limit it should reduce to the CSHJE. In
\cite{1}\cite{BFM}\cite{2} it was shown that the parameter which
selects the classical phase is the Planck constant. Therefore, in
determining the structure of the $Q$ term we have to take into
account that in the classical limit \be \lim_{\hbar\to0}Q=0\ .
\label{classicoqezero}\ee

The only possibility to reach any other state ${\cal W}^v\ne0$ starting
from ${\cal W}^0$ is that it transforms with an inhomogeneous term.
Namely as ${\cal W}^0\longrightarrow {\cal W}^v(q^v)\ne0$, it follows that for
an arbitrary ${\cal W}^a$ state \be
{\cal W}^v(q^v)=(p^v|p^a){\cal W}^a(q_a)+(q_a;q^v)\ , \label{azzoyyyxxaa10bbbb}\ee
and by (\ref{yyyxxaa10bbbb}) \be
Q^v(q^v)=(p^v|p^a)Q^a(q_a)-(q_a;q^v)\ . \label{azzo2yyyxxaa10bbbb}\ee
Let us stress that the purely quantum origin of the inhomogeneous
term $(q_a;q^v)$ is particularly transparent once one consider the
compatibility between the classical limit (\ref{classicoqezero})
and the transformation properties of $Q$ in
Eq.(\ref{azzo2yyyxxaa10bbbb}).

The ${\cal W}^0$ state plays a special role. Actually, setting
${\cal W}^a={\cal W}^0$ in Eq.(\ref{azzoyyyxxaa10bbbb}) yields $$
{\cal W}^v(q^v)=(q_0;q^v)\ , $$ so that, according to the EP
(\ref{equivalence}), all the states correspond to the
inhomogeneous part in the transformation of the ${\cal W}^0$ state
induced by some $v$--map.

Let us denote by $a,b,c,\ldots$ different $v$--transformations.
Comparing \be {\cal W}^b(q_b)=(p^b|p^a){\cal W}^a(q_a)+(q_a;q_b)=(q_0;q_b)\ ,
\label{ganzate}\ee with the same formula with $q_a$ and $q_b$
interchanged we have \be (q_b;q_a)=-(p^a|p^b)(q_a;q_b)\ ,
\label{inparticolare}\ee in particular $(q;q)=0$ More generally,
imposing the commutative diagram of maps $$
\begin{array}{c} {} \\ {} \\ A\end{array}
\begin{array}{c} {} \\ \nearrow \\ {} \end{array}
\begin{array}{c} B \\ {} \\ \longrightarrow \end{array}
\begin{array}{c} {} \\ \searrow \\ {} \end{array}
\begin{array}{c} {} \\ {} \\ C\end{array}
$$ that is comparing $$ {\cal W}^b(q_b)=(p^b|p^c){\cal W}^c(q_c)+(q_c;q_b)=
(p^b|p^a){\cal W}^a(q_a)+ (p^b|p^c)(q_a;q_c)+(q_c;q_b)\ , $$ with
(\ref{ganzate}), we obtain the basic cocycle condition \be
(q_a;q_c)=(p^c|p^b)\left[(q_a;q_b)+(q_b;q_c)\right]\ ,
\label{cociclo3}\ee which expresses the essence of the EP. In the
one dimensional case we have \be
(q_a;q_c)=\left(\partial_{q_c}q_b\right)^2(q_a;q_b)+(q_b;q_c)\ .
\label{inhomtrans}\ee It is well--known that this is satisfied by
the Schwarzian derivative. However, it turns out that it is
essentially the unique solution. More precisely, we have \cite{1}

\vspace{.333cm}

\noindent {\bf Theorem 1.} {\it Eq.}(\ref{inhomtrans}) {\it
defines the Schwarzian derivative up to a multiplicative constant
and a coboundary term.}

\vspace{.333cm}

\noindent Since the differential equation for $\mathcal{S}_0$ should
depend only on $\partial_q^k\mathcal{S}_0$, $k\geq1$, it follows
that the coboundary term must be zero, so that \cite{1} $$
(q_a;q_b)=-{\beta^2\over4m}\{q_a,q_b\}\ , $$ where
$\{f(q),q\}=f'''/f'-3(f''/f')^2/2$ is the Schwarzian derivative and
$\beta$ is a nonvanishing constant that we identify with $\hbar$. As
a consequence, $\mathcal{S}_0$ satisfies the Quantum Stationary
Hamilton--Jacobi Equation (QSHJE) \cite{1} \be {1\over
2m}\left({\partial\mathcal{S}_0(q)\over\partial q}\right)^2+V(q)-E
+{\hbar^2\over4m}\{\mathcal{S}_0,q\}=0\ . \label{1Q}\ee Note that
$\psi={\mathcal{S}_0'}^{-1/2}\left(A
e^{-{i\over\hbar}\mathcal{S}_0}+Be^{{i\over\hbar}\mathcal{S}_0}\right)$
solves the Schr\"odinger Equation (SE) \be
\left(-{\hbar^2\over2m}{\partial^2\over\partial
q^2}+V\right)\psi=E\psi\ .\label{yz1xxxx4}\ee The ratio
$w=\psi^D/\psi$, where $\psi^D$ and $\psi$ are two real linearly
independent solutions of (\ref{yz1xxxx4}) is, in deep analogy with
uniformization theory, the {\it trivializing map} transforming any
$\mathcal{W}$ to $\mathcal{W}^0\equiv0$
\cite{1}\cite{Matone:1993tj}. This formulation, proposed in
collaboration with Faraggi, extends to higher dimension and to the
relativistic case as well \cite{1}\cite{BFM}.

Let $q_{-/+}$ be the lowest/highest $q$ for which $\mathcal{W}(q)$
changes sign, we have \cite{1}

\vspace{.333cm}

\noindent {\bf Theorem 2.} {\it If} \be
V(q)-E\geq\left\{\begin{array}{ll}P_-^2
>0\ ,&q<q_-\ ,\\ P_+^2 >0\ ,&q> q_+\ ,\end{array}\right.
\label{perintroasintoticopiumeno}\ee {\it then $w$ is a local
self--homeomorphism of $\hat{\RR}={\RR}\cup\{\infty\}$ if and only
if Eq.}(\ref{yz1xxxx4}) {\it has an $L^2(\RR)$ solution.}

\vspace{.333cm}

\noindent The crucial consequence is that since the QSHJE is defined
if and only if $w$ is a local self--homeomorphism of $\hat{\RR}$, it
follows that the QSHJE by itself implies energy quantization. We
stress that this result is obtained without any probabilistic
interpretation of the wave function.

\section{Proof of Theorem 1}

The main steps in proving theorem 1 are two lemmas \cite{1}. Let
us start observing that if the cocycle condition
(\ref{inhomtrans}) is satisfied by $(f(q);q)$, then this is still
satisfied by adding a coboundary term \be
(f(q);q)\longrightarrow(f(q);q)+(\partial_qf)^2G(f(q))-G(q)\ .
\label{coboundary}\ee Since $(Aq;q)$ evaluated at $q=0$ is independent
of $A$, we have \be 0=(q;q)=(q;q)_{|q=0}=(Aq;q)_{|q=0}\ .
\label{valezero}\ee Therefore, if both $(f(q);q)$ and
(\ref{coboundary}) satisfy (\ref{inhomtrans}), then $G(0)=0$,
which is the unique condition that $G$ should satisfy. We now use
(\ref{traslazione}) to fix the ambiguity (\ref{coboundary}). First
of all observe that the differential equation we are looking for
is \be (q_0;q)={\cal W}(q)\ . \label{equazionewelook}\ee Then, recalling
that $q_0={\cal S}_0^{0^{\;-1}}\circ{\cal S}_0(q)$, we see that a necessary
condition to satisfy (\ref{traslazione}) is that $(q_0;q)$ depends
only on the first and higher derivatives of $q_0$. This in turn
implies that for any constant $B$ we have $(q_a+B;q_b) =(q_a;q_b)$
that, together with (\ref{inparticolare}), gives \be
(q_a+B;q_b)=(q_a;q_b)=(q_a;q_b+B)\ . \label{1XuBtR}\ee Let $A$ be a
non--vanishing constant and set $h(A,q)=(Aq;q)$. By (\ref{1XuBtR})
we have $h(A,q+B)=h(A,q)$, that is $h(A,q)$ is independent of $q$.
On the other hand, by (\ref{valezero}) $h(A,0)=0$ that, together
with (\ref{inparticolare}), implies \be (Aq;q)=0=(q;Aq)\ .
\label{pt11}\ee Eq.(\ref{inhomtrans}) implies $(q_a;
Aq_b)=A^{-2}((q_a;q_b)-(Aq_b;q_b))$, so that by (\ref{pt11}) \be
(q_a;Aq_b)=A^{-2}(q_a;q_b)\ . \label{1Xw}\ee By (\ref{inparticolare})
and (\ref{1Xw}) we have $$(Aq_a;q_b)=-A^{-2}(
\partial_{q_b}q_a)^2(q_b;Aq_a)
=-(\partial_{q_b}q_a)^2(q_b;q_a)= (q_a;q_b)\ ,$$ that is \be
(Aq_a;q_b)=(q_a;q_b)\ . \label{1Xu}\ee Setting $f(q)=q^{-2}(q;q^{-1})$
and noticing that by (\ref{inparticolare}) and (\ref{1Xu})
$f(Aq)=-f(q^{-1})$, we obtain \be (q;q^{-1})=0=(q^{-1};q)\ .
\label{kenzoexenasarco}\ee Furthermore, since by (\ref{inhomtrans})
and (\ref{kenzoexenasarco}) one has
$(q_a;q_b^{-1})=q_b^4(q_a;q_b)$, it follows that
$$(q_a^{-1};q_b)
=-\left(\partial_{q_b}q_a^{-1}\right)^2(q_b; q_a^{-1})
=-\left(\partial_{q_b}{q_a}\right)^2(q_b;q_a)=(q_a;q_b)\ ,
$$
so that \be (q_a^{-1};q_b)=(q_a;q_b)=q_b^{-4}(q_a;q_b^{-1})\ .
\label{pt22}\ee Since translations, dilatations and inversion are the
generators of the M\"obius group, it follows by
(\ref{1XuBtR})(\ref{1Xw})(\ref{1Xu}) and (\ref{pt22}) that

\vspace{.333cm}

\noindent {{\bf Lemma 1.}\it  Up to a coboundary term,
Eq.}(\ref{inhomtrans}) {\it implies} $$
(\gamma(q_a);q_b)=(q_a;q_b)\ , $$ $$
(q_a;\gamma(q_b))=\left(\partial_{q_b}\gamma(q_b)\right)^{-2}(q_a;q_b)\
, $$ where $\gamma(q)$ is an arbitrary $PSL(2,\CC)$
transformation.

\vspace{.333cm}

\noindent
Now observe that since $(q_a;q_b)$ should depend only on
$\partial_{q_b}^kq_a$, $k\geq1$, we have \be (q+\epsilon
f(q);q)=c_1\epsilon f^{(k)}(q)+\mathcal{O}(\epsilon^2)\ ,
\label{preliminare1}\ee where $q_a=q+\epsilon f(q)$, $q\equiv q_b$ and
$f^{(k)}\equiv\partial_q^kf$, $k\geq1$. Note that by lemma 1 and
(\ref{preliminare1}) $$(Aq+\epsilon Af(q);Aq)$$
\be =(q+\epsilon
f(q);Aq)=A^{-2}(q+\epsilon f(q);q)=A^{-2}c_1\epsilon
f^{(k)}(q)+\mathcal{O}(\epsilon^2)\ ,\label{preliminare2}\ee on the
other hand, setting $F(Aq)=Af(q)$, by (\ref{preliminare1})
$$(Aq+\epsilon Af(q);Aq)$$
$$=(Aq+\epsilon F(Aq);Aq)
=c_1\epsilon\partial_{Aq}^kF(Aq)+\mathcal{O}(\epsilon^2)=
A^{1-k}c_1\epsilon f^{(k)}(q)+\mathcal{O}(\epsilon^2)\
,$$ that compared with (\ref{preliminare2}) gives $k=3$. The above
scaling property generalizes to higher order contributions in
$\epsilon$. In particular, at order $\epsilon^n$ the quantity
$(Aq+\epsilon Af(q);Aq)$ is a sum of terms of the form
$$c_{i_1\ldots i_n}\partial_{Aq}^{i_1}\epsilon F(Aq)\cdot\cdot\cdot
\partial_{Aq}^{i_n}\epsilon F(Aq)
=c_{i_1\ldots i_n}\epsilon^nA^{n-\sum i_k}
f^{(i_1)}(q)\cdot\cdot\cdot f^{(i_n)}(q)\ ,$$ and by
(\ref{preliminare2}) $\sum_{k=1}^ni_k=n+2$. On the other hand,
since $(q_a;q_b)$ depends only on $\partial_{q_b}^kq_a$, $k\geq1$,
we have
$$i_k\geq 1\ ,\qquad k\in[1,n]\ ,$$ so that either
$$i_k=3\ ,\qquad i_j=1\ ,\qquad
j\in[1,n]\ ,\qquad j\ne k\ ,$$ or $$i_k=i_j=2\ ,\qquad i_l=1\
,\qquad l\in[1,n]\ , \qquad l\ne k,\,l\ne j\ .$$ Hence
\be(q+\epsilon f(q);q)=
\sum_{n=1}^\infty\epsilon^n\left(c_nf^{(3)}
f^{(1)^{n-1}}+d_nf^{(2)^2}f^{(1)^{n-2}}\right)\ ,\qquad d_1=0\ .
\label{preliminare4}\ee Let us now consider the transformations $$
q_b=v^{ba}(q_a)\ , \,\qquad q_c=v^{cb}(q_b)=
v^{cb}\circ v^{ba}(q_a)\ , \,\qquad
q_c=v^{ca}(q_a)\ .$$ Note that $v^{ab}=v^{ba^{-1}}$, and \be
v^{ca}=v^{cb}\circ v^{ba}\ . \label{vcacbba}\ee We can express these
transformations in the form $$ q_b=q_a+\epsilon^{ba}(q_a)\ ,$$ \be
q_c=q_b+\epsilon^{cb}(q_b)=q_b+
\epsilon^{cb}(q_a+\epsilon^{ba}(q_a))\ ,\label{epsilonabcvx}\ee
$$ q_c=q_a+\epsilon^{ca}(q_a)\ .$$  Since
$q_b=q_a-\epsilon^{ab}(q_b)$, we have $q_b=q_a-\epsilon^{ab}(q_a+
\epsilon^{ba}(q_a))$ that compared with
$q_b=q_a+\epsilon^{ba}(q_a)$ yields $$
\epsilon^{ba}+\epsilon^{ab}\circ({\bf 1}+\epsilon^{ba})=0\ ,
$$ where ${\bf 1}$ denotes the identity map. More generally,
Eq.(\ref{epsilonabcvx}) gives$$ \epsilon^{ca}(q_a)=
\epsilon^{cb}(q_b)+\epsilon^{ba}(q_a)=
\epsilon^{cb}(q_b)-\epsilon^{ab}(q_b)\ , $$ so that we obtain
(\ref{vcacbba}) with $v^{yx}={\bf 1}+\epsilon^{yx}$ \be
\epsilon^{ca}=\epsilon^{cb}\circ({\bf
1}+\epsilon^{ba})+\epsilon^{ba}= ({\bf 1}+\epsilon^{cb})\circ({\bf
1}+\epsilon^{ba})-{\bf 1}\ . \label{unpoininoFBT2}\ee Let us consider
the case in which $\epsilon^{yx}(q_x)=\epsilon f_{yx} (q_x)$, with
$\epsilon$ infinitesimal. At first--order in $\epsilon$
Eq.(\ref{unpoininoFBT2}) reads \be
\epsilon^{ca}=\epsilon^{cb}+\epsilon^{ba}\ , \label{unpoininoFBT3}\ee
in particular, $\epsilon^{ab}=-\epsilon^{ba}$. Since
$(q_a;q_b)=c_1 {\epsilon^{ab}}'''(q_b)+{\cal O}^{ab}(\epsilon^2)$, where
$'$ denotes the derivative with respect to the argument, we can
use the cocycle condition (\ref{inhomtrans}) to get $$
c_1{\epsilon^{ac}}'''(q_c)+{\cal O}^{ac}(\epsilon^2)$$
\be =(1+{\epsilon^{bc}}'(q_c))^2\left(c_1{\epsilon^{ab}}'''(q_b)+{\cal O}^{ab}(\epsilon^2)
-c_1{\epsilon^{cb}}'''(q_b)-{\cal O}^{cb}(\epsilon^2)\right)\ ,
\label{unpoininoFBT4}\ee that at first--order in $\epsilon$
corresponds to (\ref{unpoininoFBT3}). We see that $c_1\ne0$. For,
if $c_1=0$, then by (\ref{unpoininoFBT4}), at second--order in
$\epsilon$ one would have \be
{\cal O}^{ac}(\epsilon^2)={\cal O}^{ab}(\epsilon^2)-{\cal O}^{cb}(\epsilon^2)\ ,
\label{unpoininoFBT5}\ee which contradicts (\ref{unpoininoFBT3}). In
fact, by (\ref{preliminare4}) we have $$
{\cal O}^{ab}(\epsilon^2)=c_2{\epsilon^{ab}}'''(q_b){\epsilon^{ab}}'(q_b)+
d_2{{\epsilon^{ab}}''}^2(q_b)+{\cal O}^{ab}(\epsilon^3)\ ,
$$ that together with (\ref{unpoininoFBT5})
provides a relation which cannot be consistent with
$\epsilon^{ac}(q_c)=\epsilon^{ab}(q_b)-\epsilon^{cb} (q_b)$. A
possibility is that $(q_a;q_b)=0$. However, this is ruled out by
the EP, so that $$ c_1\ne0\ .$$ Higher--order contributions due to
a non--vanishing $c_1$ are obtained by using
$$q_c=q_b+\epsilon^{cb}(q_b)\ , \qquad\quad \epsilon^{ac}
(q_c)=\epsilon^{ab}(q_b)-\epsilon^{cb}(q_b)\ ,$$ and
$\epsilon^{bc}(q_c)=- \epsilon^{cb}(q_b)$ in
$c_1\partial_{q_c}^3\epsilon^{ac}(q_c)$ and in
$$c_1\left(2\partial_{q_c}\epsilon^{bc}(q_c)+{\partial_{q_c}\epsilon^{bc}
(q_c)}^2\right)\partial^3_{q_b}\left({\epsilon^{ab}}(q_b)-{\epsilon^{cb}}
(q_b)\right)\ .$$ Note that one can also consider the case in
which both the first-- and second--order contributions to
$(q_a;q_b)$ are vanishing. However, this possibility is ruled out
by a similar analysis. In general, one has that if the first
non--vanishing contribution to $(q_a;q_b)$ is of order
$\epsilon^n$, $n\geq2$, then, unless $(q_a;q_b)=0$, the cocycle
condition (\ref{inhomtrans}) cannot be consistent with the
linearity of (\ref{unpoininoFBT3}). Observe that we proved that
$c_1\ne0$ is a necessary condition for the existence of solutions
$(q_a;q_b)$ of the cocycle condition (\ref{inhomtrans}), depending
only on the first and higher derivatives of $q_a$. Existence of
solutions follows from the fact that the Schwarzian derivative
$\{q_a,q_b\}$ solves (\ref{inhomtrans}) and depends only on the
first and higher derivatives of $q_a$.

The fact that $c_1=0$ implies $(q_a;q_b)=0$, can be also seen by
explicitly evaluating the coefficients $c_n$ and $d_n$. These can
be obtained using the same procedure considered above to prove
that $c_1\ne0$. Namely, inserting the expansion
(\ref{preliminare4}) in (\ref{inhomtrans}) and using $q_c=q_b+
\epsilon^{cb}(q_b)$,
$\epsilon^{ac}(q_c)=\epsilon^{ab}(q_b)-\epsilon^{cb} (q_b)$ and
$\epsilon^{bc}(q_c)=-\epsilon^{cb}(q_b)$, we obtain \be
c_n=(-1)^{n-1}c_1\ ,\qquad d_n={3\over2}(-1)^{n-1}(n-1)c_1\ ,
\label{cenneedennexf}\ee which in fact are the coefficients one
obtains expanding $c_1\{q+\epsilon f(q),q\}$. However, we now use
only the fact that $c_1\ne0$, as the relation $(q+\epsilon
f(q);q)=c_1\{q+\epsilon f(q),q\}$ can be proved without making the
calculations leading to (\ref{cenneedennexf}). Summarizing, we
have

\vspace{.333cm}

\noindent {{\bf Lemma 2.}\it  If $$q_a=q_b+\epsilon^{ab}(q_b)\ ,$$
the unique solution of Eq.}(\ref{inhomtrans}){\it, depending only
on the first and higher derivatives of $q_a$, is}
$$
(q_a;q_b)=c_1{\epsilon^{ab}}'''(q_b)+\mathcal{O}^{ab}(\epsilon^2)\
,\qquad c_1\ne0\ . $$

\noindent

\vspace{.333cm}

It is now easy to prove that, up to a multiplicative constant and
a coboundary term, the Schwarzian derivative is the unique
solution of the cocycle condition (\ref{inhomtrans}). Let us first
note that
$$
[q_a;q_b]=(q_a;q_b)-c_1\{q_a;q_b\}\ ,
$$
satisfies the cocycle condition
$$
[q_a;q_c]=\left({\partial_{q_c}q_b}\right)^2\left([q_a;q_b]-[q_c;q_b]
\right)\ .
$$
In particular, since both $(q_a;q_b)$ and $\{q_a;q_b\}$ depend
only on the first and higher derivatives of $q_a$, we have, as in
the case of $(q+\epsilon f(q);q)$, that
$$
[q+\epsilon f(q);q]=\tilde c_1\epsilon f^{(3)}(q)+{\cal O}(\epsilon^2)\
,
$$
where either $\tilde c_1\ne0$ or $[q+\epsilon f(q);q]=0$. However,
since $\{q+\epsilon f(q);q\}=\epsilon f^{(3)}(q)+{\cal O}(\epsilon^2)$
and $(q+\epsilon f(q);q)=\epsilon f^{(3)}(q)+{\cal O}(\epsilon^2)$, we
have $\tilde c_1=0$ and the Lemma yields $[q+\epsilon f(q);q]=0$.
Therefore, we have that the EP univocally implies that
$$
(q_a;q_b)=-{\beta^2\over4m}\{q_a,q_b\}\ ,
$$
where for convenience we replaced $c_1$ by $-\beta^2/4m$. This
concludes the proof of theorem 1.

We observe that despite some claims \cite{Ovsienko1}, we
have not be able to find in the literature a complete and close
proof of the above theorem (see also \cite{DMS}). We thank D.B. Fuchs for a
bibliographic comment concerning the above theorem.

In deriving the equivalence of states we considered the case of
one--particle states with identical masses. The generalization to
the case with different masses is straightforward. In particular,
the right hand side of Eq.(\ref{inhomtrans}) gets multiplied by
$m_b/m_a$, so that the cocycle condition becomes
$$
m_a(q_a;q_c)=m_a\left(\partial_{q_c}q_b\right)^2(q_a;q_b)+m_b(q_b;q_c)\
, $$ explicitly showing that the mass appears in the denominator
and that it refers to the label in the first entry of
$(\cdot\,;\cdot)$, that is \be (q_a;q_b)=-{\hbar^2\over
4m_a}\{q_a;q_b\}\ . \label{bella}\ee The QSHJE (\ref{1Q}) follows
almost immediately by (\ref{bella}) \cite{1}.

The above investigation may be applied to CFT. Let us consider a
local conformal transformation of the stress tensor in a 2D CFT.
The infinitesimal variation of $T$ is given by \be \delta_\epsilon
T(w)=-{1\over12}c\partial_w^3\epsilon(w)-2T(w)\partial_w\epsilon(w)-\epsilon(w)\partial_wT(w)
\ , \label{hJ1}\ee where $c$ is the central charge. The finite version
of such a transformation is \be \tilde T(w)=(\partial_w
z)^2T(z)+{c\over12}\{w,z\} \ . \label{hJ2}\ee While it is immediate to
see that (\ref{hJ2}) implies (\ref{hJ1}), the viceversa is not
evident. A possible way to prove (\ref{hJ2}) is just to set \be
\tilde T(w)=(\partial_w z)^2T(z)+k(w;z) \ , \label{hJ3}\ee and then to
impose the cocycle condition which will show that $(w;z)$ is
proportional to $\{w,z\}$.  Comparison with the infinitesimal
transformation (\ref{hJ1}) fixes the constant $k$.

In \cite{BFM} it has been shown that the cocycle condition fixes
the higher dimensional version of the Schwarzian derivative. In
this respect we observe that its definition seems an open question
in mathematical literature. While in the one dimensional case the
QSHJE reduces to a unique differential equation, this is not
immediate in the higher dimensional case. However, it turns out
that such a reduction exists upon introducing an antisymmetric
tensor \cite{BFM} (in this respect it is worth noticing that some
author introduces a connection to define the higher dimensional
Schwarzian derivative).

A basic feature of the cocycle condition is that it implies, as it
should, the higher dimensional M\"obious invariance with respect
to $q_a$ in $(q_a;q_b)$ (with similar properties with respect to
$q_b$). In particular, in \cite{BFM} it has been shown that \be
(q^a;q^b)=-{\hbar^2\over2m}\left[(p^b|p^a) {\Delta^aR^a\over
R^a}-{\Delta^bR^b\over R^b}\right]\ . \label{cocicloide}\ee It would
be interesting to consider such a definition in the context of the
transformation properties of the stress tensor in higher
dimensional CFTs.

\section{Proof of Theorem 2}

The QSHJE is equivalent to \be \{w,q\}=-{4m\over\hbar^2}{\cal W}(q)\
, \label{cosicchesothat}\ee where $w=\psi^D/\psi$ with $\psi^D$ and
$\psi$ two real linearly independent solutions of the Schr\"odinger
equation. Existence of this equation requires some conditions on the
continuity properties of $w$ and its derivatives. Since the QSHJE is
the consequence of the EP, we can say that the EP imposes some
constraints on $w=\psi^D/\psi$. These constraints are nothing but
the existence of the QSHJE (\ref{1Q}) or, equivalently, of
Eq.(\ref{cosicchesothat}). That is, implementation of the EP imposes
that $\{w,q\}$ exists, so that \be w\ne cnst,\;w\in C^2(\RR)
\;and\;\partial_q^2w\; differentiable \;on\;\RR\ .
\label{ccnndraft}\ee These conditions are not complete. The reason
is that, as we have seen, the implementation of the EP requires that
the properties of the Schwarzian derivative be satisfied. Actually,
its very properties, derived from the EP, led to the identification
$(q_a;q_b)=-\hbar^2 \{q_a,q_b\}/4m$. Therefore, in order to
implement the EP, the transformation properties of the Schwarzian
derivative and its symmetries must be satisfied. In deriving the
transformation properties of $(q_a;q_b)$ we noticed how, besides
dilatations and translations, there is a highly non--trivial
symmetry such as that under inversion. Therefore, we have that
(\ref{cosicchesothat}) must be equivalent to
$$
\{w^{-1},q\}=-{4m\over\hbar^2}{\cal W}(q)\ .
$$
A property of the Schwarzian derivative is duality between its
entries \be \{w,q\}=-\left({\partial w\over\partial
q}\right)^2\{q,w\}\ . \label{pksxjq}\ee This shows that the invariance
under inversion of $w$ reflects in the invariance, up to a
Jacobian factor, under inversion of $q$. That is
$\{w,q^{-1}\}=q^4\{w,q\}$, so that the QSHJE
(\ref{cosicchesothat}) can be written in the equivalent form \be
\{w,q^{-1}\}=-{4m\over\hbar^2}q^4{\cal W}(q)\ . \label{oidjwI939}\ee In
other words, starting from the EP one can arrive to either
Eq.(\ref{cosicchesothat}) or Eq.(\ref{oidjwI939}). The consequence
of this fact is that since under
$$
q\rightarrow {1\over q}\ ,
$$
$0^{\pm}$ maps to $\pm\infty$, we have to extend (\ref{ccnndraft})
to the point at infinity. In other words, (\ref{ccnndraft}) should
hold on the extended real line $\hat\RR =\RR\cup\{\infty\}$. This
aspect is related to the fact that the M\"obius transformations,
under which the Schwarzian derivative transforms as a quadratic
differential, map circles to circles. We stress that we are
considering the systems defined on $\RR$ and not $\hat\RR$. What
happens is that the existence of the QSHJE forces us to impose
smoothly joining conditions even at $\pm\infty$, that is
(\ref{ccnndraft}) must be extended to \be w\ne cnst,\;w\in
C^2(\hat\RR)\;and\;\partial_q^2w\;differentiable\;on\;\hat\RR\ .
\label{ccnn}\ee One may easily check that $w$ is a M\"obius
transformation of the trivializing map \cite{1}. Therefore,
Eq.(\ref{pksxjq}), which is defined if and only if $w(q)$ can be
inverted, that is if $\partial_q w\ne 0$, $\forall q\in\RR$, is a
consequence of the cocycle condition (\ref{cociclo3}). By
(\ref{oidjwI939}) we see that also local univalence should be
extended to $\hat\RR$. This implies the following joining condition
at spatial infinity \be w(-\infty)=\left\{\begin{array}{ll}
w(+\infty)\ , & {\rm for}\quad w(-\infty)\ne\pm \infty\ ,\\
-w(+\infty)\ ,& {\rm for}\quad w(-\infty)=\pm\infty\
.\end{array}\right. \label{specificandoccnn}\ee As illustrated by
the non--univalent function $w=q^2$, the apparently natural choice
$w(-\infty)=w(+\infty)$, one would consider also in the $w(-\infty)=
\pm\infty$ case, does not satisfy local univalence.

We saw that the EP implied the QSHJE (\ref{1Q}). However, although
this equation implies the SE, we saw that there are aspects
concerning the canonical variables which arise in considering the
QSHJE rather than the SE. In this respect a natural question is
whether the basic facts of QM also arise in our formulation. A
basic point concerns a property of many physical systems such as
energy quantization. This is a matter of fact beyond any
interpretational aspect of QM. Then, as we used the EP to get the
QSHJE, it is important to understand how energy quantization
arises in our approach. According to the EP, the QSHJE contains
all the possible information on a given system. Then, the QSHJE
itself should be sufficient to recover the energy quantization
including its structure. In the usual approach the quantization of
the spectrum arises from the basic condition that in the case in
which $\lim_{q\to\pm\infty}{\cal W}>0$, the wave--function should vanish
at infinity. Once the possible solutions are selected, one also
imposes the continuity conditions whose role in determining the
possible spectrum is particularly transparent in the case of
discontinuous potentials. For example, in the case of the
potential well, besides the restriction on the spectrum due to the
$L^2(\RR)$ condition for the wave--function (a consequence of the
probabilistic interpretation of the wave--function), the spectrum
is further restricted by the smoothly joining conditions. Since
the SE contains the term $\partial_q^2\psi$, the continuity
conditions correspond to an existence condition for this equation.
On the other hand, also in this case, the physical reason
underlying this request is the interpretation of the
wave--function in terms of probability amplitude. Actually,
strictly speaking, the continuity conditions come from the
continuity of the probability density $\rho=|\psi|^2$. This
density should also satisfy the continuity equation
$\partial_t\rho+\partial_qj=0$, where $j=i\hbar(\psi
\partial_q\bar\psi-\bar\psi\partial_q\psi)/2m$. Since for stationary states
$\partial_t\rho=0$, it follows that in this case $j=cnst$.
Therefore, in the usual formulation, it is just the interpretation
of the wave--function in terms of probability amplitude, with the
consequent meaning of $\rho$ and $j$, which provides the physical
motivation for imposing the continuity of the wave--function and
of its first derivative.

Now observe that in our formulation the continuity conditions arise
from the QSHJE. In fact, (\ref{ccnn}) implies continuity of
$\psi^D$, $\psi$, with $\partial_q\psi^D$ and $\partial_q\psi$
differentiable, that is \be
EP\;\rightarrow\;(\psi^D,\psi)\;continuous\; and
\;(\psi^{D'},\psi')\;differentiable\ .
\label{equivalenzaederivata}\ee

In the following we will see that if $V(q)>E$, $\forall q\in\RR$,
then there are no solutions such that the ratio of two real
linearly independent solutions of the SE corresponds to a local
self--homeomorphism of $\hat\RR$. The fact that this is an
unphysical situation can be also seen from the fact that the case
$V>E$, $\forall q\in\RR$, has no classical limit. Therefore, if
$V>E$ both at $-\infty$ and $+\infty$, a physical situation
requires that there are at least two points where $V-E=0$. More
generally, if the potential is not continuous, $V(q)-E$ should
have at least two turning points. Let us denote by $q_-$ ($q_+$)
the lowest (highest) turning point. Note that by
(\ref{perintroasintoticopiumeno}) we have
$$
\int^{-\infty}_{q_-}dx\kappa(x)=-\infty\
,\qquad\quad\int^{+\infty}_{q_+}dx\kappa(x)=+ \infty\ ,
$$
where $\kappa=\sqrt{2m(V-E)}/\hbar$. Before going further, let us
stress that what we actually need to prove is that, in the case
(\ref{perintroasintoticopiumeno}), the joining condition
(\ref{specificandoccnn}) requires that the corresponding SE has an
$L^2(\RR)$ solution. Observe that while (\ref{ccnn}), which
however follows from the EP, can be recognized as the standard
condition (\ref{equivalenzaederivata}), the other condition
(\ref{specificandoccnn}), which still follows from the existence
of the QSHJE, and therefore from the EP, is not directly
recognized in the standard formulation. Since this leads to energy
quantization, while in the usual approach one needs one more
assumption, we see that there is a quite fundamental difference
between the QSHJE and the SE. We stress that (\ref{ccnn}) and
(\ref{specificandoccnn}) guarantee that $w$ is a local
self--homeomorphism of $\hat\RR$.

Let us first show that the request that the corresponding SE has
an $L^2(\RR)$ solution is a sufficient condition for $w$ to
satisfy (\ref{specificandoccnn}). Let $\psi\in L^2(\RR)$ and
denote by $\psi^D$ a linearly independent solution. As we will
see, the fact that $\psi^D\not\propto\psi$ implies that if
$\psi\in L^2(\RR)$, then $\psi^D\notin L^2(\RR)$. In particular,
$\psi^D$ is divergent both at $q=-\infty$ and $q=+\infty$. Let us
consider the real ratio
$$
w={A\psi^D+B\psi\over C\psi^D+D\psi}\ ,
$$
where $AD-BC\ne 0$. Since $\psi\in L^2(\RR)$, we have \be
\lim_{q\rightarrow\pm\infty}w=\lim_{q\rightarrow\pm\infty}
{A\psi^D+B\psi\over C\psi^D+D\psi}={A\over C}\ , \label{arbitygvxy}\ee
that is $w(-\infty)=w(+\infty)$. In the case in which $C=0$ we
have
$$
\lim_{q\rightarrow\pm\infty}w=\lim_{q\rightarrow\pm\infty}{A\psi^D
\over D\psi}=\pm\epsilon\cdot\infty\ ,
$$
where $\epsilon=\pm1$. The fact that ${A\psi^D/D\psi}$ diverges for
$q\to\pm\infty$ follows from the mentioned properties of $\psi^D$
and $\psi$. It remains to check that if
$\lim_{q\to-\infty}{A\psi^D/D\psi}=-\infty$, then
$\lim_{q\to+\infty}{A\psi^D/D\psi}=+\infty$, and vice versa. This
can be seen by observing that
$$
\psi^D(q)=c\psi(q)\int^q_{q_0}dx\psi^{-2}(x)+d\psi(q)\ ,
$$
$c\in\RR\backslash\{0\}$, $d\in\RR$. Since $\psi\in L^2(\RR)$ we
have $\psi^{-1}\not\in L^2(\RR)$ and
$\int^{+\infty}_{q_0}dx\psi^{-2}=+\infty$,
$\int^{-\infty}_{q_0}dx\psi^{-2}=-\infty$, implying that
$\psi^D(-\infty)/\psi
(-\infty)=-\epsilon\cdot\infty=-\psi^D(+\infty)/\psi(+\infty)$,
where $\epsilon={\rm sgn}\,c$.

We now show that the existence of an $L^2(\RR)$ solution of the SE
is a necessary condition to satisfy the joining condition
(\ref{specificandoccnn}). We give two different proofs of this, one
is based on the WKB approximation while the other one uses Wronskian
arguments. In the WKB approximation, we have \be
\psi={A_-\over\sqrt{\kappa}}e^{-\int^q_{q_-}dx\kappa}
+{B_-\over\sqrt{\kappa}}e^{\int^q_{q_-}dx\kappa},\quad q\ll q_-\ ,
\label{Pantani1}\ee and \be
\psi={A_+\over\sqrt{\kappa}}e^{-\int^q_{q_+}dx\kappa}
+{B_+\over\sqrt{\kappa}}e^{\int^q_{q_+}dx\kappa},\quad q\gg q_+\ .
\label{Pantani2}\ee In the same approximation, a linearly
independent solution has the form
$$
\psi^D={A_-^D\over\sqrt{\kappa}}e^{-\int^q_{q_-}dx\kappa}
+{B_-^D\over {\kappa}}e^{\int^q_{q_-}dx\kappa},\quad q\ll q_-\ .
$$
Similarly, in the $q\gg q_+$ region we have
$$
\psi^D={A_+^D\over\sqrt{\kappa}}e^{-\int^q_{q_+}dx\kappa}
+{B_+^D\over\sqrt{\kappa}}e^{\int^q_{q_+}dx\kappa},\quad q\gg q_+\ .
$$
Note that (\ref{Pantani1}) and (\ref{Pantani2}) are derived by
solving the differential equations corresponding to the WKB
approximation for $q\ll q_-$ and $q\gg q_+$, so that the
coefficients of $\kappa^{-1/2}\exp\pm\int^q_{q_-}dx \kappa$, {\it
e.g.} $A_-$ and $B_-$ in (\ref{Pantani1}), cannot be simultaneously
vanishing. In particular, the fact that $\psi^D\not\propto\psi$
yields \be A_-B_-^D-A_-^DB_-\ne 0\ ,\qquad A_+B_+^D-A_+^DB_+\ne 0\ .
\label{PantaniGirodItaliaeTour}\ee Let us now consider the case in
which, for a given $E$ satisfying (\ref{perintroasintoticopiumeno}),
any solution of the corresponding SE diverges at least at one of the
two spatial infinities, that is \be \lim_{q\rightarrow +\infty}
(|\psi(-q)|+|\psi(q)|)=+\infty\ . \label{caruccioe}\ee This implies
that there is a solution diverging both at $q=-\infty$ and $q=+
\infty$. In fact, if two solutions $\psi_1$ and $\psi_2$ satisfy
$\psi_1(- \infty)=\pm\infty$, $\psi_1(+\infty)\ne\pm\infty$ and
$\psi_2(-\infty)\ne\pm \infty$, $\psi_2(+\infty)=\pm\infty$, then
$\psi_1+\psi_2$ diverges at $\pm \infty$. On the other hand,
(\ref{PantaniGirodItaliaeTour}) rules out the case in which all the
solutions in their WKB approximation are divergent only at one of
the two spatial infinities, say $-\infty$. Since, in the case
(\ref{perintroasintoticopiumeno}), a solution which diverges in the
WKB approximation is itself divergent (and vice versa), we have that
in the case (\ref{perintroasintoticopiumeno}), the fact that all the
solutions of the SE diverge only at one of the two spatial
infinities cannot occur.

Let us denote by $\psi$ a solution which is divergent both at
$-\infty$ and $+\infty$. In the WKB approximation this means that
both $A_-$ and $B_+$ are non--vanishing, so that
$$
\psi{}_{\;\stackrel{\sim}{q\rightarrow-\infty}\;}{A_-\over\sqrt\kappa}e^{-
\int^q_{q_-}dx\kappa},\quad\qquad
\psi{}_{\;\stackrel{\sim}{q\rightarrow+\infty}
\;}{B_+\over\sqrt{\kappa}}e^{\int^q_{q_+} dx\kappa}\ .
$$
The asymptotic behavior of the ratio $\psi^D/\psi$ is given by
$$
\lim_{q\rightarrow-\infty}{\psi^D\over\psi}={A_-^D\over A_-}\
,\qquad\quad
\lim_{q\rightarrow+\infty}{\psi^D\over\psi}={B_+^D\over B_+}\ .
$$
Note that since in the case at hand any divergent solution also
diverges in the WKB approximation, we have that (\ref{caruccioe})
rules out the case $A^D_-= B_+^D=0$. Let us then suppose that
either $A_-^D=0$ or $B_+^D=0$. If $A_-^D=0$, then $w(-\infty)=0\ne
w(+\infty)$. Similarly, if $B_+^D=0$, then $w(+\infty)=0 \ne
w(-\infty)$. Hence, in this case $w$, and therefore the
trivializing map, cannot satisfy (\ref{specificandoccnn}). On the
other hand, also in the case in which both $A_-^D$ and $B_+^D$ are
non--vanishing, $w$ cannot satisfy Eq.(\ref{specificandoccnn}).
For, if $A_-^D/A_-=B_+^D/B_+$, then
$$
\phi=\psi-{A_-\over A^D_-}\psi^D=\psi-{B_+\over B^D_+}\psi^D\ ,
$$
would be a solution of the SE whose WKB approximation has the form
$$
\phi={B_-\over\sqrt{\kappa}}e^{\int^q_{q_-}dx\kappa},\qquad q\ll
q_-\ ,
$$
and
$$
\phi={A_+\over\sqrt{\kappa}}e^{-\int^q_{q_+}dx\kappa},\qquad q\gg
q_+\ .
$$
Hence, if $A_-^D/A_-=B_+^D/B_+$, then there is a solution whose
WKB approximation vanishes both at $-\infty$ and $+\infty$. On the
other hand, we are considering the values of $E$ satisfying
Eq.(\ref{perintroasintoticopiumeno}) and for which any solution of
the SE has the property (\ref{caruccioe}). This implies that no
solutions can vanish both at $-\infty$ and $+\infty$ in the WKB
approximation. Hence
$$
{A_-^D\over A_-}\ne{B_+^D\over B_+}\ ,
$$
so that $w(-\infty)\ne w(+\infty)$. We also note that not even the
case $w(- \infty)=\pm\infty=-w(+\infty)$ can occur, as this would
imply that $A_-=B_+=0$, which in turn would imply, against the
hypothesis, that there are solutions vanishing at $q=\pm\infty$.
Hence, if for a given $E$ satisfying
(\ref{perintroasintoticopiumeno}), any solution of the
corresponding SE diverges at least at one of the two spatial
infinities, we have that the trivializing map has a discontinuity
at $q=\pm \infty$. As a consequence, the EP cannot be implemented
in this case so that this value $E$ cannot belong to the physical
spectrum.

Therefore, the physical values of $E$ satisfying
(\ref{perintroasintoticopiumeno}) are those for which there are
solutions which are divergent neither at $-\infty$ nor at $+\infty$.
On the other hand, from the WKB approximation and
(\ref{perintroasintoticopiumeno}), it follows that the
non--divergent solutions must vanish both at $-\infty$ and
$+\infty$. It follows that the only energy levels satisfying the
property (\ref{perintroasintoticopiumeno}), which are compatible
with the EP, are those for which there exists the solution vanishing
both at $\pm\infty$. On the other hand, solutions vanishing as
$\kappa^{-1/2}\exp\int^q_{q_-}dx \kappa$ at $-\infty$ and
$\kappa^{-1/2}\exp-\int^q_{q_+}dx\kappa$ at $+\infty$, with
$P^2_\pm>0$, cannot contribute with an infinite value to
$\int^{+\infty}_{-\infty}dx\psi^2$. The reason is that existence of
the QSHJE requires that $\{e^{{2i\over\hbar}{\cal S}_0},q\}$ be
defined and this, in turn, implies that any solution of the SE must
be continuous. On the other hand, since $\psi$ is continuous, and
therefore finite also at finite values of $q$, we have
$\int^{q_b}_{q_a}dx\psi^2<+\infty$ for all finite $q_a$ and $q_b$.
In other words, the only possibility for a continuous function to
have a divergent value of $\int^{+ \infty}_{-\infty}dx\psi^2$ comes
from its behavior at $\pm\infty$. Therefore, since the
implementation of the EP in the case
(\ref{perintroasintoticopiumeno}) requires that the corresponding
$E$ should admit a solution with the behavior
$$
\psi{}_{\;\stackrel{\sim}{q\rightarrow-\infty}\;}{A_-\over\sqrt{\kappa}}
e^{\int^q_{q_-}dx\kappa},\qquad\psi{}_{\;\stackrel{\sim}{q\rightarrow+
\infty}\;}{B_+\over\sqrt{\kappa}}e^{-\int^q_{q_+}dx\kappa}\ ,
$$
we have the following basic fact

\vspace{.333cm}

\noindent {\it The values of $E$ satisfying \be
V(q)-E\geq\left\{\begin{array}{ll}P_-^2>0\ ,&q<q_-\ ,\\
P_+^2>0\ ,&q>q_+\ ,
\end{array}\right.
\label{asintoticopiumenofgt}\ee are physically admissible if and only
if the corresponding SE has an $L^2(\RR)$ solution. }

\vspace{.333cm}

We now give another proof of the fact that if ${\cal W}$ is of the type
(\ref{asintoticopiumenofgt}), then the corresponding SE must have
an $L^2(\RR)$ solution in order to satisfy
(\ref{specificandoccnn}). In particular, we will show that this is
a necessary condition. That this is sufficient has been already
proved above.

By Wronskian arguments, which can be found in Messiah's book
\cite{Messiah}, imply that if $V(q)-E\geq P_+^2>0$, $q>q_+$, then
as $q\rightarrow +\infty$, we have ($P_+>0$)

\begin{itemize}
\item[{\bf --}]{There is a solution of the SE that vanishes at least as
$e^{-P_+q}$.}
\item[{\bf --}]{Any other linearly independent solution diverges at least as
$e^{P_+q}$.}
\end{itemize}

\noindent Similarly, if $V(q)-E\geq P_-^2>0$, $q<q_-$, then as
$q\rightarrow- \infty$, we have ($P_->0$)

\begin{itemize}
\item[{\bf --}]{There is a solution of the SE that vanishes at least as
$e^{P_-q}$.}
\item[{\bf --}]{Any other linearly independent solution diverges at least as
$e^{-P_-q}$.}
\end{itemize}

\noindent These properties imply that if there is a solution of
the SE in $L^2(\RR)$, then any solution is either in $L^2(\RR)$ or
diverges both at $-\infty$ and $+\infty$. Let us show that the
possibility that a solution vanishes only at one of the two
spatial infinities is ruled out. Suppose that, besides the
$L^2(\RR)$ solution, which we denote by $\psi_1$, there is a
solution $\psi_2$ which is divergent only at $+\infty$. On the
other hand, the above properties show that there exists also a
solution $\psi_3$ which is divergent at $-\infty$. Since the
number of linearly independent solutions of the SE is two, we have
$\psi_3=A \psi_1+B\psi_2$. However, since $\psi_1$ vanishes both
at $-\infty$ and $+\infty$, we see that $\psi_3=A\psi_1+B\psi_2$
can be satisfied only if $\psi_2$ and $\psi_3$ are divergent both
at $-\infty$ and $+\infty$. This fact and the above properties
imply that

\vspace{.333cm}

\noindent {\it If the SE has an $L^2(\RR)$ solution, then any
solution has two possible asymptotics}

\begin{itemize}
\item[{\bf --}]{Vanishes both at $-\infty$ and $+\infty$ at least as $e^{P_-q}$
and $e^{-P_+q}$ respectively.}
\item[{\bf --}]{Diverges both at $-\infty$ and $+\infty$ at least as $e^{-P_-q}$
and $e^{P_+q}$ respectively.}
\end{itemize}

\vspace{.333cm}

\noindent Similarly, we have

\vspace{.333cm}

\noindent {\it If the SE does not admit an $L^2(\RR)$ solution,
then any solution has three possible asymptotics}

\begin{itemize}
\item[{\bf --}]{Diverges both at $-\infty$ and $+\infty$ at least as $e^{-P_-q}$
and $e^{P_+q}$ respectively.}
\item[{\bf --}]{Diverges at $-\infty$ at least as $e^{-P_-q}$ and vanishes at
$+\infty$ at least as $e^{-P_+q}$.}
\item[{\bf --}]{Vanishes at $-\infty$ at least as $e^{P_-q}$ and diverges at
$+\infty$ at least as $e^{P_+q}$.}
\end{itemize}

\vspace{.333cm}

\noindent Let us consider the ratio $w=\psi^D/\psi$ in the latter
case. Since any different choice of linearly independent solutions
of the SE corresponds to a M\"obius transformation of $w$, we can
choose
$$
\psi^D_{\;\stackrel{\sim}{q\rightarrow-\infty}\;}a_-e^{P_-q}\
,\qquad\qquad
\psi^D_{\;\stackrel{\sim}{q\rightarrow+\infty}\;}a_+e^{P_+q}\ ,
$$
and
$$
\psi_{\;\stackrel{\sim}{q\rightarrow-\infty}\;}b_-e^{-P_-q}\
,\qquad\qquad
\psi_{\;\stackrel{\sim}{q\rightarrow+\infty}\;}b_+e^{-P_+q}\ ,
$$
were by $\sim$ we mean that $\psi^D$ and $\psi$ either diverge or
vanish ``at least as". Their ratio has the asymptotic
$$
{\psi^D\over\psi}{}_{\;\stackrel{\sim}{q\rightarrow-\infty}\;}c_-e^{2P_-q}
\rightarrow0\ ,\quad{\psi^D\over\psi}{}_{\;\stackrel{\sim}{q
\rightarrow+\infty}\;}c_+e^{2P_+q}\rightarrow\pm\infty\ ,
$$
so that $w$ cannot satisfy Eq.(\ref{specificandoccnn}). This
concludes the alternative proof of the fact that, in the case
(\ref{asintoticopiumenofgt}), the existence of the $L^2(\RR)$
solution is a necessary condition in order
(\ref{specificandoccnn}) be satisfied. The fact that this is a
sufficient condition has been proved previously in deriving
Eq.(\ref{arbitygvxy}).

The above results imply that the usual quantized spectrum arises
as a consequence of the EP.

Let us note that we are considering real solutions of the SE.
Thus, apparently, in requiring the existence of an $L^2(\RR)$
solution, one should specify the existence of a real $L^2(\RR)$
solution. However, if there is an $L^2(\RR)$ solution $\psi$, this
is unique up to a constant, and since also $\bar\psi\in L^2(\RR)$
solves the SE, we have that an $L^2(\RR)$ solution of the SE is
real up to a phase.

\section{The two--particle model}

The EP leads to the introduction of length scales
\cite{1}\cite{BFM}\cite{2}, a fact related to the nontriviality of
the quantum potential, even in the case of $\mathcal{W}^0$. We
note that also $\mathcal{S}_0$, as follows by the EP, is never
trivial, in particular \be\mathcal{S}_0\ne cnst\ ,\qquad
\forall\mathcal{W}\in{{\cal H}}\ . \label{assolutamentebasilare}\ee

The QSHJE (\ref{1Q}), first investigated by Floyd in a series of
important papers \cite{Floyd}, has been studied and reviewed by
several authors \cite{vari}\cite{reviews}.

The real solution of the QSHJE (\ref{1Q}) is $$
e^{{2i\over\hbar}\mathcal{S}_0\{\delta\}}=e^{i\alpha}
{w+i\bar\ell\over w-i\ell}\ ,\qquad \delta\equiv\{\alpha,\ell\}\
,$$ with $\alpha\in{\RR}$ and $\ell=\ell_1+i\ell_2$,
$\ell_1\neq0$, are integration constants. The condition
$\ell_1\neq0$ is necessary for $\mathcal{S}_0$ and the quantum
potential $Q$ to be well--defined.

The formulation has a manifest duality between real pairs of
linearly independent solutions \cite{1}, a property strictly
related to Legendre duality \cite{140} (see \cite{seealso} for
related issues). A similar structure also appears in
uniformization theory  of Riemann surfaces
\cite{1}\cite{Matone:1993tj}. Whereas in the standard approach one
usually considers only one solution of the SE, {\it i.e.} the
wave--function itself, in our formulation the relevant formulas
contain the linear combination $\psi^D+i\ell \psi$. Since
$\ell_1\neq0$, $\psi^D$ and $\psi$ appears always in pair. So,
Legendre duality, nontriviality of $\mathcal{S}_0$ and $Q$ are
deeply related features which are direct consequences of the EP.
In turn, these properties imply the appearance of fundamental
constants such as the Planck length \cite{1}\cite{BFM}\cite{2}.
The simplest way to see this is to consider the SE in the trivial
case, that is $\partial_{q_0}^2\psi=0$, so that $\psi^D=q_0$,
$\psi=1$ and the typical combination reads $q_0+i\ell_0$, implying
that $\ell_0\equiv\ell$ should have the dimension of a length. The
fact that $\ell$ has the dimension of a length is true for any
state. Since $\ell_0$ appears in the QSHJE with
$\mathcal{W}^0\equiv0$, the system does not provide any
dimensional quantity, so that we have to introduce some
fundamental length. The appearance of fundamental constants also
arises in considering the limits $\hbar\to0$ and $E\to0$ in the
case of a free particle \cite{1}. So, for example, a consistent
expression for the quantum potential associated to the trivial
state $\mathcal{W}^0$, which vanishes as $\hbar\to0$, is $$
Q^0={\hbar\over 4m}\{\mathcal{S}_0^0,q_0\}=-{\hbar^3 G
\over2mc^3}{1\over|q_0-i\lambda_p|^4}\ ,$$ where
$$\lambda_p=\sqrt{\hbar G/c^3}\ ,$$ is the Planck length. However,
in considering the classical limit of the reduced action one
should include the gravitational contribution. So, for example, it
is clear that also at the classical level the reduced action for a
pair of ``free" particles should include the Newton potential.
This may be related to the above mentioned appearance of
fundamental constants in the QSHJE \cite{1}\cite{BFM}\cite{2}.
Related to this is the Floyd observation that in the classical
limit there is a residual indeterminacy depending on the
integration constants \cite{Floyd}. Thus we see that the classical
limit may in fact lead to some effect which is of quantum origin
even if $\hbar$ does not appear explicitly. We also note that, in
principle, the Planck constant may appear in macroscopic
phenomena. This indicates that it is worth studying the structure
of the quantum potential also at large scales.

It seems that the fundamental properties of $Q$ have not yet fully
been investigated because the usual solutions one finds are
essentially trivial. This is due to a clearly unsatisfactory
identification, that may lead to some inconsistency, of
$Re^{{i\over\hbar}\mathcal{S}_0}$ with the wave--function. As
noticed by Floyd \cite{Floyd}, if
$$Re^{{i\over\hbar}\mathcal{S}_0}\ ,$$ solves the SE, then the
wave--function has the general form
$R(Ae^{-{i\over\hbar}\mathcal{S}_0}+Be^{{i\over\hbar}\mathcal{S}_0})$.
This simple remark has important consequences. So, for example, note
that a real wave--function, such as the one for bound states, simply
implies $|A|=|B|$ rather than $\mathcal{S}_0=0$. As Einstein noticed
in a letter to Bohm, the latter would imply that a quantum particle
in a box is at rest and starts moving in the classical limit.
Therefore, besides the mathematical consistency, identifying the
wave--function with $Re^{{i\over\hbar}\mathcal{S}_0}$ cannot in
general lead to a quantum analog of the reduced action. This change
in the definition of $\mathcal{S}_0$ implies a new view of $Q$ which
needs to be further investigated. In this respect we note that $Q$
provides particle's response to an external perturbation. For
example, in the case of tunnelling, where according to the standard
definition $\mathcal{S}_0$ would be vanishing, the attractive nature
of $Q$ guarantees the reality of the conjugate momentum \cite{1}. As
a consequence, the role of this intrinsic energy, which is a
property of all forms of matter, should manifest itself through
effective interactions depending on the above fundamental constants.

It is therefore natural to consider the so called {\it
two--particle model} \cite{2}, consisting of two free particles in
the three dimensional space. This provides a simple physical model
to investigate the structure of the interaction provided by $Q$.
The QSHJE decomposes in equations for the center of mass and for
the relative motion. The latter is the QSHJE $$ {1\over
2m}(\nabla\mathcal{S}_0)^2-E-{\hbar^2\over 2m}{\Delta R\over R}=0\
, \quad \nabla \cdot(R^2\nabla\mathcal{S}_0)=0\ ,$$ where
$$r=r_1-r_2\ , \qquad m={m_1m_2 \over m_1+m_2}\ .$$ Due to the
quantum potential, the QSHJE has solutions in which the relative
motion is not free as in the classical case. Also note that the
quantum potential is negative definite. Since
$\psi=Re^{{i\over\hbar}\mathcal{S}_0}$ is a solution of the SE, we
have $$\mathcal{S}_0={\hbar\over 2i}\ln(\psi/\overline\psi)\ ,$$
so that $$ (\nabla\mathcal{S}_0)^2=-{\hbar^2\over
4|\psi|^4}(\overline \psi\nabla\psi -\psi\nabla\overline \psi)^2\
,$$ where
$$\psi=\sum_{l=0}^\infty\sum_{m=-l}^l\sum_{j=1}^2 c_{lmj}
R_{klj}(r)Y_{lm}(\theta,\phi)\ ,$$ with $Y_{lm}$ the spherical
harmonics and $R_{klj}$ the solutions of the radial part of the SE
\cite{2}.

As $m=l\to\infty$ $P^l_l\propto\sin^l\theta$ vanishes unless
$\theta=\pi/2$, and the motion is on a plane as in the classical
orbits. However, since $\lim_{l\to\infty}\partial_\theta
P^l_l(\cos\theta)=0$, we have $$\lim_{m\sim
l\to\infty}\partial_\theta P^m_l(\cos\theta)=0\ .$$ Thus,
considering solutions with $c_{lmj}\neq0$ only for sufficiently
large $m$ and $l$, we have $$\nabla\psi=
\sum_{\{lmj\}}\left(c_{lmj}R_{klj}'Y_{lm}, 0, {i\over
r}c_{lmj}R_{klj}mY_{lm}\right)\ .$$

Depending on the coefficients $c_{lmj}$,
$(\nabla\mathcal{S}_0)^2/2m$ may contain nontrivial terms which do
not cancel as $\hbar\to0$. These may arise as a deformation of the
classical kinetic term, which includes the centrifugal term. The
$c_{lmj}$, which may depend on fundamental constants, fix the
structure of the possible interaction in the two--particle model.
The $c_{lmj}$ may be related to some boundary conditions implied
by the geometry and the matter content of the three--dimensional
space. This would relate the fundamental constants to possible
collective effects \cite{2} which may depend on cosmological
aspects.

\section{Acknowledgments}

I am grateful to Hans--Thomas Elze and to the sponsors of the
workshop ``Dice 2004" for the invitation to talk in Piombino. Work
partially supported by the European Community's Human Potential
Programme under contract HPRN-CT-2000-00131 Quantum Spacetime and
by the INTAS Research Project N.2000-254.

\bibliography{apssamp}

\end{document}